\documentclass{article}

\usepackage{arxiv}

\usepackage[utf8]{inputenc} 
\usepackage[T1]{fontenc}    
\usepackage{hyperref}       
\usepackage{url}            
\usepackage{booktabs}       
\usepackage{amsfonts}       
\usepackage{nicefrac}       
\usepackage{microtype}      
\usepackage{lipsum}
\usepackage{graphicx}
\usepackage{amsmath,amssymb}

\newcommand{\mi}{\mathrm{i}}
\graphicspath{ {./images/} }

\title{Full-spectrum high resolution modeling of the dielectric function of water}

\author{J.~Fiedler\thanks{Centre for Materials Science and Nanotechnology, Department of Physics, University of Oslo, P. O. Box 1048 Blindern, NO-0316 Oslo, Norway}\\
Physikalisches Institut\\Albert-Ludwigs-Universit{\"a}t Freiburg\\Hermann-Herder-Str. 3\\79104 Freiburg, Germany\\
\texttt{johannes.fiedler@physik.uni-freiburg.de}
\And
M.~Bostr{\"o}m\footnotemark[1]\\
Department of Energy and Process Engineering\\Norwegian University of Science and Technology\\NO-7491 Trondheim, Norway\\\texttt{mathias.a.bostrom@ntnu.no}
\And
C.~Persson\\Centre for Materials Science and Nanotechnology\\Department of Physics\\University of Oslo\\P. O. Box 1048 Blindern\\NO-0316 Oslo, Norway
 \And
 I.~Brevik\\
Department of Energy and Process Engineering\\Norwegian University of Science and Technology\\NO-7491 Trondheim, Norway
\And
R.~Corkery\thanks{Applied Physical Chemistry, KTH Royal Institute of Technology, SE 100 44 Stockholm, Sweden}\\
Surface and Corrosion Science\\Department of Chemistry\\KTH Royal Institute of Technology\\SE 100 44 Stockholm, Sweden
\And
S.~Y.~Buhmann\\
Physikalisches Institut\\Albert-Ludwigs-Universit{\"a}t Freiburg\\Hermann-Herder-Str. 3\\79104 Freiburg, Germany
\And
D.~F.~Parsons\\School of Engineering and IT\\Murdoch University\\90 South St\\Murdoch, WA 6150, Australia\\
\texttt{d.parsons@murdoch.edu.au}
}

\begin{document}
\maketitle
\begin{abstract}
In view of the vital role of water in chemical and physical processes, an exact knowledge of its dielectric function over a large frequency range is important.  In this article we report on  currently available measurements of the dielectric function of water at room temperature (25$^{\circ}$C) across the full electromagnetic spectrum: microwave, IR, UV and X-ray (up to 100 eV).   We provide parameterisations of the complex dielectric function of water with two Debye (microwave) oscillators and high resolution of IR and UV/X-ray oscillators. We also report dielectric parameters for ice-cold water with a microwave/IR spectrum measured  at $0.4^\circ$C, while taking the UV spectrum from 25$^{\circ}$C (assuming negligible temperature dependence in UV).  We illustrate the consequences of the model via calculations of van der Waals interactions of  gas molecules near water surfaces, and an assessment of the thickness of water films on ice and ice films on water. In contrast to earlier models of ice-cold water, we predict that a micron-scale layer of ice is stabilised on a bulk water surface. Similarly, the van der Waals interaction promotes complete freezing rather than supporting a thin premelting layer of water on a bulk ice surface.  Density-based extrapolation from warm to cold water of the dielectric function at imaginary frequencies is found to be satisfactory in the microwave but poor (40\% error) at IR frequencies.
\end{abstract}


\section{Introduction}

 A large variety  of biological, chemical and environmental processes take place in water~\cite{doi:10.1021/jp806376e}. The huge amount of water  at the earth's surface supports a multitude of dissolved gases and nano-particles that are significant in climate change; cf.,  for instance,  black carbon~\cite{Lund2018}. Further issues related to water are the behaviour of oil films~\cite{Chen2018}, the synthesis of nano-materials for medical applications~\cite{Tsukiashi2018,Lee2019}, the behaviour of  two-dimensional materials~\cite{Mehmood2018},  and the behaviour of colloidal suspensions~\cite{Lan2018,Michael}.  The functional mechanism of membranes~\cite{Ries2019,You2019} is also a relevant class. The description of all these systems requires good knowledge of the dielectric function of water on the real frequency axis as well as, for calculation of van der Waals interactions, imaginary frequencies. An insufficient  knowledge of these quantities can result in drastically wrong predictions. Typically, there are two different approaches: either performing measurements, making use of the Kramers--Kronig relationship as a helpful ingredient, or using quantum mechanical simulations. In modern works, material properties are obtained from a mixture of both methods aiming to get accurate results. However, while water is structurally one of the simplest molecules it is one of the most difficult when it comes to simulating its bulk properties~\cite{GUILLOT2002219}. Thus, the properties of pure water are hard to estimate. Applications where water is included, such as biological or chemical reactions, often require a  quantum description~\cite{SokhanJonesCipciganCrainMartyna2015,Rick2016}  to obtain a  reliable molecular simulation. In practice, when considering   processes involving  chemical and biological reactions in the presence of water,  the influence of water   on intermolecular forces can at best be estimated. Investigations have shown a dependence on dispersion forces in the self-assembling structure of molecules in free-space~\cite{doi:10.1021/jacs.7b07884}, as well as   on a surface~\cite{doi:10.1021/acsnano.7b05204}. 
 
 Further, dispersion forces in liquids play an important role in the self-arrangement of amorphous structures. When  particles are brought into a solution and heated,  the solvent may evaporate,   increasing the particle density. Due to  van der Waals interactions between the particles, they stick together and can build an amorphous solid~\cite{Velikov2335}. The properties of the solvent have a very large influence on the resulting structure. Although  dispersion forces are universal in nature and present  between all bodies and particles,   their magnitude cannot be so easily manipulated as is usually the case for  electrical forces \cite{PhysRevLett.106.064501}.  For some scenarios a rough approximation of the functional dependence is sufficient,  and detailed  spectroscopic effects are not essential.
 
Aqueous media such as atmospheric water droplets and water reservoirs  (e.g. lakes, oceans, and frozen soil) may act as sources and sinks for greenhouse gases. Fluxes of carbon dioxide and methane from wetlands in north America and their potential dependence on  temperature have  recently been studied experimentally~\cite{Pugh}. Emission of dissolved greenhouse gas molecules from a water surface can influence the environment on  both a global and a local scale.

We have collected a wide range of optical data  (up to 100 eV) from the literature ranging across the full electromagnetic spectrum including microwave, IR, UV and X-ray. We make use of these very detailed optical data to prepare improved parameterisations for the dielectric functions room temperature water (25$^\circ$C). We also present a parameterisation  for ice-cold water ($0.4^\circ$C) using detailed microwave and IR data. For ice-cold water, we assume the UV/X-ray spectrum does not change significantly between 25$^\circ$C and $0.4^\circ$C (detailed measured data is not available). However we present a independent parameterisation of the UV spectrum of ice-cold in order to allow for causality consistency at the two temperatures.  We compare the influence of our dielectric functions  with some models from the literature on the dispersion force between a dissolved  particle in water and the water surface. Surprisingly, we find strong differences for the dispersion forces with water interface, varying from attraction to repulsion for molecules near a water surface  depending on one's  choice of dielectric function. Furthermore for ice--water systems, a previous model for ice-cold water \cite{Elbaum273Kwater} leads to the prediction of a thin premelting layer on bulk ice surfaces while Lifshitz forces evaluated with our new model for ice cold water supports  a micron-sized ice layer on a water surface. These observed sensitive behaviours can be used to judge which of these dielectric function models  is the more accurate one.

\section{Available data and models for dielectric function of water}

Water is one of the most complicated systems to be studied theoretically. A complete model describing its properties, especially its dielectric properties, does not exist. In order to be able to perform calculations with this specific material, experimental data are required covering the complete spectral range. In this work, we used the optical data for microwave scattering from CRC Handbook of optical constants~\cite{CRC18}, Kaatze \textit{el al.}~\cite{Kaatze89} and Buchner \textit{et al.}~\cite{BUCHNER199957}; the infrared data from Hale \textit{et al.}~\cite{Hale:73} and Irvine \textit{et al.}~\cite{IRVINE1968324}; the visible range from Bertie \textit{et al.}~\cite{Bertie96} and Pinkley~\cite{Pinkley:77}; and the ultraviolet data from Hayashi \textit{et al.}~\cite{Hayashi15}. In order to combine all these data sets, the dielectric function is typically fitted to a damped harmonic oscillator model
\begin{equation}
    \varepsilon(\omega) = 1+\sum_i \frac{c_i}{1-\mi \omega \tau_i} +\sum_j \frac{c_j \omega_j^2}{\omega_j^2-\mi\omega\gamma_j-\omega^2} \, ,\label{eq:model}
\end{equation}
with the oscillator strengths $c_j$ which directly correspond to the plasma frequencies, the resonance frequencies $\omega_j$, the damping constants $\gamma_j$ together with  Debye damping terms with  peak height $c_i$ and  relaxation time $\tau_i$.

Previous models, in particular models from Parsegian \& Weiss~\cite{PARSEGIAN1981285}, Roth \& Lenhoff~\cite{ROTH1996637}, Elbaum \& Schick~\cite{Elbaum273Kwater} and Dagastine \textit{et al.}~\cite{Dagastine298Kwater}, used a similar fitting method combining all data available at the time. Here, we  compare our results against these as well as the newest data sets of Wang \textit{et al.}~\cite{WANG201754}, where only the ultraviolet data~\cite{Hayashi15} are used. We acknowledge a correction required in Wang \textit{et al.}, which should have $f_3 =8.31\,(\rm{eV})^2$ in Tbl.~5 in Ref.~\cite{WANG201754}. The differences between these models can be partly understood from their chronological order. The first model was published by Parsegian \& Weiss, who started with an 11-oscillator fit. Some years later, Roth \& Lenhoff recognised the impact of higher spectral ranges and repeated the fit with an extension to higher frequencies. Later, Elbaum \& Schick found  better agreement by not considering the Debye oscillators and correcting the static dielectric constant afterwards by hand. Tbl.~\ref{tbl:overview} summarises these points.

\begin{table}\centering
\begin{tabular}{|c|cccc|}
\hline
Model & MW & IR & UV & Comment\\\hline
Parsegian \& Weiss & Ref.~\cite{TF9534901003} & Ref.~\cite{IRVINE1968324} & Ref.~\cite{Heller74} & Fixing refractive index~\cite{PARSEGIAN1981285} \\
Roth \& Lenhoff & Ref.~\cite{TF9534901003} & Ref.~\cite{IRVINE1968324} &Ref.~\cite{Heller74} & Range extended to higher frequencies \\
Elbaum \& Schick & - & Ref.~\cite{IRVINE1968324} & Ref.~\cite{Heller74} & Fixing zero-frequency value~\cite{Buckley58}\\\hline
\end{tabular}
\caption{Overview about the early models for the dielectric function of water.} \label{tbl:overview}
\end{table}

The point of departure for this work is the publication by Dagastine \textit{et al.} \cite{Dagastine298Kwater} from 2000, which was based on a variety of measurements: the microwave spectra from  Schwan, Kaatze and others \cite{Schwan76,Kaatze89,CRC18}, the infrared spectrum from Hale \cite{Hale:73}, the visible spectrum from Bertie \cite{Bertie96} and the ultraviolet spectrum from Hayashi \cite{Hayashi98}. Unfortunately, a fitting was not performed. We have also used this collection of data sets, updating the ultraviolet range with measurements  satisfying the f-sum rule~\cite{theoryquantumliquids,PhysRevB.78.075410}, and adding further measurements closing some gaps. Dagastine \textit{et al.} studied room temperature water. We consider ice-cold water also. Here, we found a mismatch between two data sets for the ice cold water, which will be reported later.

\begin{figure}
    \centering
    \includegraphics[width=0.6\textwidth]{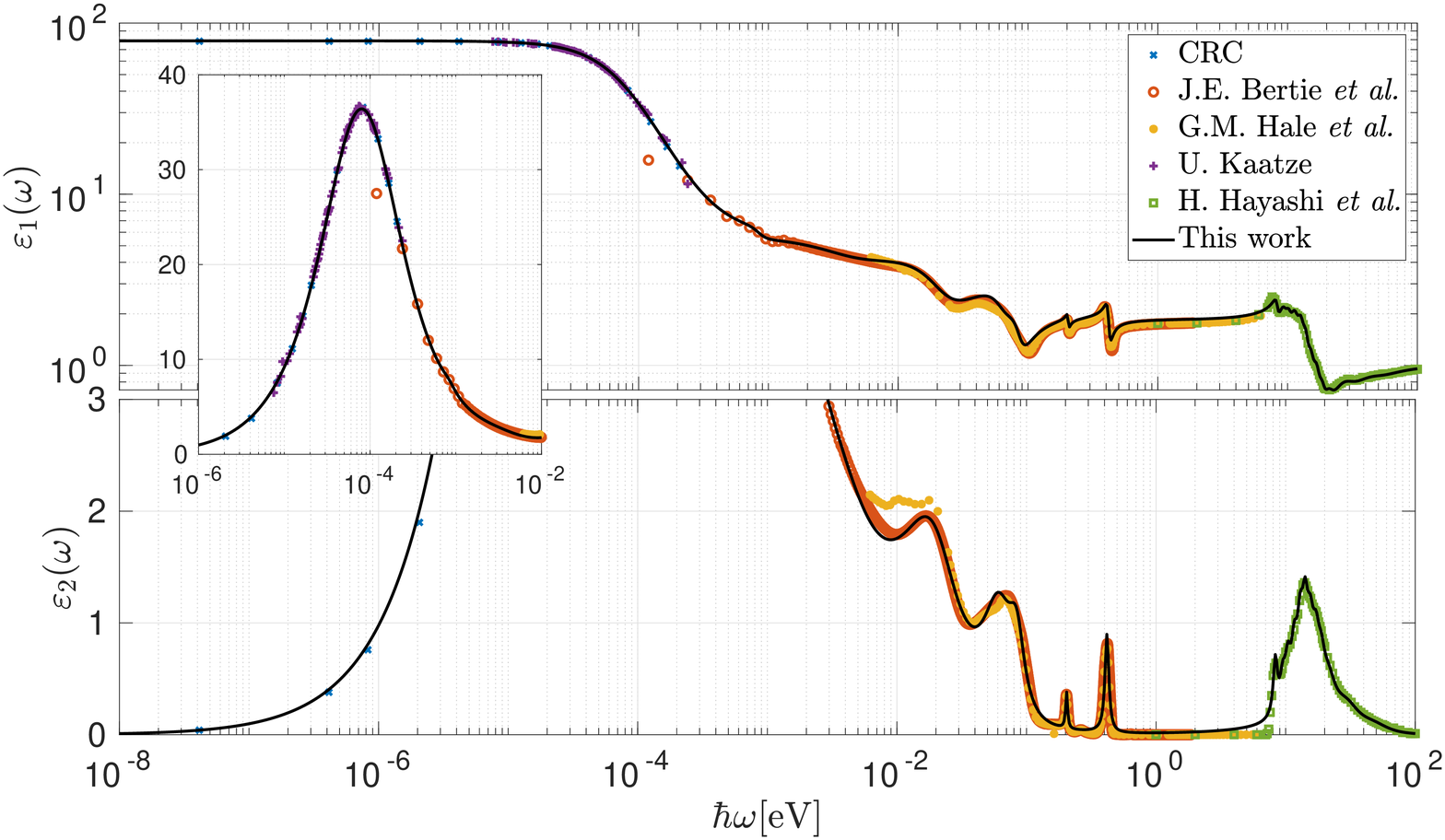}
    \caption{Dielectric spectrum of water at 25$^{\circ}$C, Shown are the real $\varepsilon_{1}(\omega)$ and imaginary $\varepsilon_{2}(\omega)$ components at real frequencies. The experimental data from CRC~\cite{CRC18} (blue crosses), Bertie \textit{et al.}~\cite{Bertie96} (red circles), Hale \textit{et al.}~\cite{Hale:73} (yellow asterisks), Hayashi \textit{et al.}~\cite{Hayashi15} (green squares), and Kaatze~\cite{Kaatze89} (violet plus signs) are shown as discrete symbols. Our fit for 25$^{\circ}$C is shown as a black solid line.}
    \label{fig:roomtemp}
\end{figure}

For room temperature water, we combined the measurements from Refs.~\cite{CRC18,Bertie96,Hale:73,Hayashi15,Kaatze89} similar to previous works~\cite{Dagastine298Kwater} with the updated UV spectrum and  fitted the experimental data to Eq.~(\ref{eq:model}). The resulting parameters are given in Table~\ref{tbl:fit25} and the curves are depicted in Fig.~\ref{fig:roomtemp}. For  ice-cold water ($T=273.55 \,\rm{K}$), we used the microwave data from Refs.~\cite{CRC18, BUCHNER199957}, for the IR and visible range~\cite{Pinkley:77,IRVINE1968324}. A measured UV spectrum for ice-cold water is not available, hence we assumed minimal temperature dependence across 0--25$^\circ$C in the UV and used the same room-temperature data from Hayashi~\cite{Hayashi15}. This assumption can be justified since thermal energy ($k_{\rm B}T$) at room temperature is 25.8 meV, far below UV energies ($> 1$ eV) reported in Ref.~\cite{Hayashi15}. That is, there is little change from 0$^\circ$C to 25$^\circ$C in the excited electron populations that generate the UV spectrum.  Parameters for ice-cold water are listed in Table~\ref{tbl:fit0}.

We provide distinct parameterisations for UV oscillators at each temperature in order to allow for causality consistency via the Kramers-Kronig relation linking real and imaginary frequencies, \cite{LandauLifshitz-ElectrodynContMedia-v8}
\begin{equation}
    \varepsilon(\mi\xi) = 1 +  \frac{2}{\pi} \int_{0}^{\infty} 
    \frac{\omega \varepsilon_2(\omega)} {\xi^2 + \omega^2} d\omega
    \label{eq:KK_imagfreq}
\end{equation}
where $\varepsilon_2(\omega)$ is the imaginary part of the dielectric function at real frequencies.  We evaluated $\varepsilon(\mi\xi)$ numerically using room-temperature data and used all 3 data sets, $\varepsilon_1(\omega)$, $\varepsilon_2(\omega)$ and $\varepsilon(\mi\xi)$ (i.e. $\varepsilon(\mi\omega)$) to fit parameter for UV oscillators (with oscillator frequency $\omega_j> 1$ eV).  We consider the numerical integration for Eq.~\ref{eq:KK_imagfreq}  to be insufficiently accurate  at lower frequencies to justify the use of  $\varepsilon(\mi\xi)$ when fitting Debye or IR oscillators. Likewise gaps in the IR data of ice-cold water preclude using a numerical   $\varepsilon(\mi\xi)$ at $0.4^\circ$C. In this way we obtain causality-consistent parameters for the UV parameters of room temperature water, while our UV parameters for ice-cold water, evaluated using  UV data borrowed from room temperature at real frequencies, may be taken as a prediction.

\section{Fitted Oscillators}

We applied the following process to determine model parameters. Firstly for room temperature water, we used the collected experimental data (real and imaginary components $\varepsilon_{1}(\omega)$  and $\varepsilon_{2}(\omega)$) at real frequencies in the Kramers--Kronig transformation of Eq.~\ref{eq:KK_imagfreq} to compute a numerical dielectric spectrum $\varepsilon(\mi\omega)$ over imaginary frequencies. We then divided the spectrum into different regimes, microwave ($< 10^{-3}$ eV), infrared (far IR $10^{-2}$--$10^{-1}$ eV, near IR 0.1--1 eV) and UV (optical/UV 1--20 eV, soft X-ray 20--100 eV). Parameters for oscillators active in each regime were fitted separately. The Parsegian--Weiss parameters \cite{PARSEGIAN1981285} were used to provide an initial guess.  IR and UV oscillator parameters were frozen in order to fit the parameters for the initial single Debye oscillator using the real and imaginary data $\varepsilon_{1}(\omega)$  and $\varepsilon_{2}(\omega)$ in the real microwave regime simultaneously (for the purposes of fitting oscillator parameters we judged the Kramers--Kronig numerical data over imaginary frequencies to be reliable  in the imaginary UV regime, not in the microwave or IR regimes). Fitting was performed using Fityk~\cite{Wojdyr2010}. To optimise the fit of the Debye oscillator, a temporary  constant (fitted to $-0.48$) was added to the real component during the initial microwave fit to account for any deviation due to the initial value of the high frequency parameters. This constant was eliminated in subsequent iterations.  Next, microwave and UV parameters were fixed while infrared oscillators were fitted against $\varepsilon_{1}(\omega)$  and $\varepsilon_{2}(\omega)$ in the real IR regime. We found the fit at room temperature could be improved with additional IR oscillators (7 IR oscillators in total), while only 5 IR oscillators could be justified for ice-cold water, a discrepancy addressed below. Adding  more oscillators  was numerically unstable, with the fitting software  generating unphysical negative values of some additional parameters. Then, with microwave and IR parameters fixed, UV parameters were fitted simultaneously against all three data sets, $\varepsilon_{1}(\omega)$  and $\varepsilon_{2}(\omega)$ in the real UV regime and $\varepsilon(\mi\omega)$ in the imaginary UV regime. A simultaneous fit over real and imaginary UV frequencies ensures that the model conforms with causality (expressed in the Kramers--Kronig relationships), and that the response at large imaginary frequencies is consistent with the numerical Kramers--Kronig data.    Peaks in the $\varepsilon_{2}(\omega)$ data were used to set initial guess of additional oscillators, finally using 11 UV oscillators.  With  oscillator parameters  refined over all frequency regimes, we repeated the procedure, now using refined values as an initial guess.

\begin{figure}
    \centering
    \includegraphics[width=0.6\textwidth]{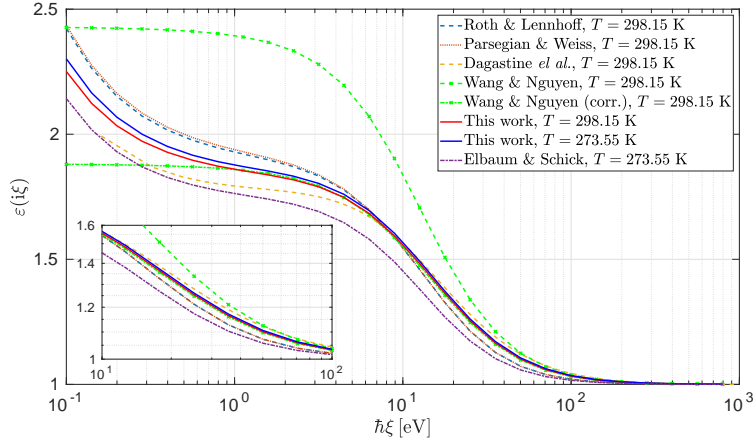}
    \caption{Dielectric function of water for imaginary frequencies.  Here "This work" is the parameterised dielectric functions at 298.15\,K and at 273.55\,K. For comparision we show the room temperature models from Parsegian and Weiss~\cite{PARSEGIAN1981285}, Roth and Lenhoff~\cite{ROTH1996637}, Dagastine \textit{et al.}~\cite{Dagastine298Kwater}, Wang and Nguyen~\cite{WANG201754} (and a corrected version of this parameterisation). We also show the ice cold water from Elbaum and Schick~\cite{Elbaum273Kwater}.}
    \label{fig:epsixi}
\end{figure}
Figure~\ref{fig:epsixi} shows the results for the different models of the dielectric function of water on the imaginary frequency axis used to evaluate van der Waals interactions (cf. Eqs~\ref{eq:Hamaker} and~\ref{eq:CasimirPolder} below). The deviation from Dagastine's model is quite low in the plotted regime, which is due to the fact that we used the same datasets in the low-frequency regime and only changed the ultraviolet data. An interesting effect can be observed for the model derived from Wang \& Nguyen which only takes into account the ultraviolet regime. It strongly overestimates the value of the dielectric function at imaginary frequencies. However, it fits well on the real axis. This is caused by the restriction to five symmetric oscillators. Detailed investigations in this regime show that the right-hand part of the tail (for frequencies larger than the resonance frequency) fits very well, but the left-hand  tail (low frequencies) is overestimated. The discrepancy is even more dramatic on the imaginary frequency axis. We found that a sufficiently good match to room temperature data with causality consistency (including numerical $\epsilon(\mi\omega)$ in UV fitting) was achieved with  a model of 7 damped IR oscillators  and 12 damped UV oscillators  (plus two Debye oscillators) given in Table~\ref{tbl:fit25}.

\begin{figure}
    \centering
    \includegraphics[width=0.6\columnwidth]{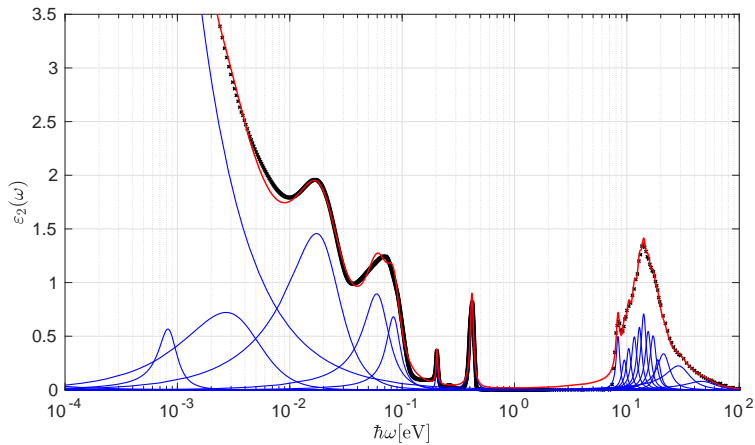}
    \caption{Water dielectric function (imaginary part $\varepsilon_{2}(\omega)$, red curve) as a function of real frequencies $\hbar\omega$, showing component oscillator peaks (blue curves) in IR and UV frequencies.  Experimental data are also plotted for comparison (black crosses).
    }
    \label{fig:eps_oscillators}
\end{figure}

The observed microwave relaxation peak was slightly asymmetric, weaker than a single Debye oscillator at frequencies lower than the peak and stronger at higher frequencies. In principle, Debye asymmetry can be captured by a Cole--Davidson model \cite{Calderwood2003,Fu2014}. But a positive asymmetry parameter (that would represent collective oscillator correlations) in predicted to generate an asymmetry which is stronger at higher frequencies, the opposite behaviour to that observed in the water data here. A negative parameter would fit well, but implies an anticorrelation of collective oscillations, which we judge to be unphysical. Instead, during the second refinement of parameters we added a second Debye oscillator which reproduces the asymmetry in the experimental data well. This conforms with the two-Debye analysis of gigahertz data performed by Buchner \textit{et al.} \cite{BUCHNER199957}.  Fitting in each frequency regime was then repeated to refine all parameters and ensure self-consistency between frequency regimes.  Fitted parameters  are collected in Table.~\ref{tbl:fit25}. We note that our data extends to X-rays up to 100 eV, where previous analyses \cite{PARSEGIAN1981285,Dagastine298Kwater} had access only to 40 eV.  The new data have enabled our fit to resolve oscillators at 21, 30 and 50 eV.  Fitted IR and UV oscillator peaks are shown against the model for $\varepsilon_{2}(\omega)$ in Fig.~\ref{fig:eps_oscillators}, showing a cluster of IR oscillators centred around 0.02 eV and a cluster of UV oscillators centred around 14 eV.

\begin{figure}
    \centering
    \includegraphics[width=0.6\textwidth]{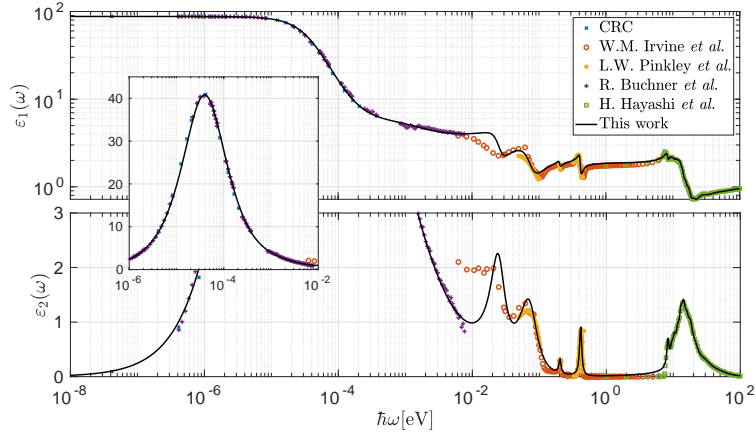}
    \caption{Dielectric spectrum of water at 0$^{\circ}$C, Shown are the real $\varepsilon_{1}(\omega)$ and imaginary $\varepsilon_{2}(\omega)$ components at real frequencies. Here "This work" is the full set of experimental data used. The experimental data from CRC~\cite{CRC18} (blue crosses), Irvine \textit{et  al.}~\cite{IRVINE1968324} (red circles), Hayashi \textit{et  al.}~\cite{Hayashi15} (green squares), Pinkley \textit{et  al.}~\cite{Pinkley:77} (yellow asterisks), and Buchner \textit{et  al.}~\cite{BUCHNER199957} (violet plus signs) are shown as discrete symbols. Our fit is shown as a black solid line.}
    \label{fig:water_icecold}
\end{figure}

The parameters fitted to experimental for ice-cold water (0.4$^\circ$C) (with  real frequency UV data borrowed from 25$^\circ$C) are given in Table~\ref{tbl:fit0}.  There is a gap in the experimental data at 0$^\circ$C in the terahertz regime (0.008--0.06 eV), such that only 5 IR oscillators could be resolved.  Our fit, shown in Fig.~\ref{fig:water_icecold}, predicts an oscillator at 0.025 eV in this experimental gap, which is somewhat at odds with Irvine's fit. However the experimental data for warm water do include terahertz data, and reveal the presence of an oscillator at 0.021 eV.  Our prediction of an oscillator in this region in ice-cold water is therefore consistent with measurements of warm water. With a numerical evaluation of $\epsilon(\mi\omega)$ not sufficiently accurate to use in a causality consistent  fitting of UV data, the real frequency UV data supported the resolution of 11 damped UV oscillators. 

Since we assumed that real frequency UV data  did not change between 25$^\circ$C and $0.4^\circ$C the frequencies and oscillator strengths of UV oscillators are identical (within the available precision) at both temperatures.  A subtle  temperature dependence in the UV oscillators was found due to our use of $\varepsilon(\mi\omega)$ at room temperature to get causality consistency: the damping coefficients of some UV oscillators at room temperature are slightly smaller than their ice-cold counterparts. Our UV parameters for ice-cold water may be considered a prediction rather than a fitted measurement.

\begin{table}\centering
\begin{tabular}{|ccc|}
\hline
$\omega_j [\rm{eV}]$ & $c_j$ & $\gamma_j [\rm{eV}^{-1}]$ \\\hline
\multicolumn{3}{|c|}{Microwave}\\
\multicolumn{3}{|c|}{$c_1 = 0.47 \pm 0.10 \, , $  $1/\tau_1 = (6.84 \pm 2.09) \times 10^{-6} \, \rm{eV}$}  \\
\multicolumn{3}{|c|}{$c_2 = 72.62 \pm 0.12 \, , $  $ 1/\tau_2 = (7.98\pm 0.025) \times 10^{-5} \, \rm{eV}$} \\
\multicolumn{3}{|c|}{Infrared}\\
$(8.46 \pm 0.28) \times 10^{-4}$   & $(2.59 \pm 0.53) \times 10^{-1}  $  & $(3.92 \pm 0.97) \times 10^{-4}  $ \\
$(4.19 \pm 0.28) \times 10^{-3}$   & $ 1.04 \pm 0.05                 $  & $(7.43 \pm 0.09) \times 10^{-3}  $\\
$(2.12 \pm 0.001) \times 10^{-2}$  & $ 1.62 \pm 0.05                 $  & $(2.60 \pm 0.05) \times 10^{-2}  $\\
$(6.25 \pm 0.04) \times 10^{-2}$   & $(5.55 \pm 0.20) \times 10^{-1}  $  & $(3.98 \pm 0.11) \times 10^{-2}  $\\
$(8.49 \pm 0.02) \times 10^{-2}$   & $(2.38 \pm 0.13) \times 10^{-1}  $  & $(2.99 \pm 0.07) \times 10^{-2}  $\\
$(2.04 \pm 0.0009) \times 10^{-1}$ & $(1.34 \pm 0.03) \times 10^{-2}  $  & $(8.43 \pm 0.26) \times 10^{-3}  $\\
$(4.18 \pm 0.0007) \times 10^{-1}$ & $(7.17 \pm 0.03) \times 10^{-2}  $  & $(3.41 \pm 0.02) \times 10^{-2}  $\\
\multicolumn{3}{|c|}{Ultraviolet}\\
$8.34  \pm 0.01$ & $(4.47 \pm 0.03) \times 10^{-2}$ & $0.75 \pm 0.04$ \\
$9.50  \pm 0.07$ & $(3.27 \pm 1.38) \times 10^{-2}$ & $1.12 \pm 0.25$\\
$10.41 \pm 0.06$ & $(4.66 \pm 2.28) \times 10^{-2}$ & $1.26 \pm 0.36$\\
$11.67 \pm 0.07$ & $(6.67 \pm 2.94) \times 10^{-2}$ & $1.58 \pm 0.40$\\
$12.95 \pm 0.09$ & $(7.42 \pm 4.58) \times 10^{-2}$ & $1.65 \pm 0.50$\\
$14.13 \pm 0.08$ & $(9.30 \pm 6.34) \times 10^{-2}$ & $1.86 \pm 0.60$\\
$15.50 \pm 0.12$ & $(7.79 \pm 6.81) \times 10^{-2}$ & $2.22 \pm 0.89$\\
$17.17 \pm 0.14$ & $(7.9 \pm 7.2) \times 10^{-2}$ & $2.7 \pm 1.1 $\\
$18.89 \pm 0.03$ & $(4.18 \pm 0.69) \times 10^{-2}$ & $2.82 \pm 0.20$\\
$21.45 \pm 0.12$ & $(1.07 \pm 0.23) \times 10^{-1}$ & $6.87 \pm 0.64$\\
$30.06 \pm 0.31$ & $(1.33 \pm 0.28) \times 10^{-1}$ & $18.28 \pm 2.13$\\
$49.45 \pm 1.41$ & $(5.66 \pm 1.10) \times 10^{-2}$ & $36.28 \pm 2.22$\\\hline

\end{tabular}
\caption{Fitting parameters for room temperature water ($T=  298.15\,\rm{K}$).}
\label{tbl:fit25}
\end{table}

\begin{table}\centering
\begin{tabular}{|ccc|}
\hline
$\omega_j [\rm{eV}]$ & $c_j$ & $\gamma_j [\rm{eV}^{-1}]$ \\\hline
\multicolumn{3}{|c|}{Microwave}\\
\multicolumn{3}{|c|}{$c_1 = 0.75 \pm 0.13 \, , $  $1/\tau_1 = 	(1.92	 \pm 0.56) \times 10^{-6} \, \rm{eV}$} \\
\multicolumn{3}{|c|}{$c_2 = 81.42 \pm 0.20 \, , $  $ 1/\tau_2 = (	3.934\pm 0.014) \times 10^{-5} \, \rm{eV}$} \\
\multicolumn{3}{|c|}{Infrared}\\
$(6.59 \pm 1.20) \times 10^{-3}$  & $1.614 \pm 0.063  $      & $0.029 \pm 0.011  $\\
$(2.502 \pm 0.086) \times 10^{-2}$  & $1.058 \pm 0.056  $    & $0.0151 \pm 0.0023 $\\
$(7.344 \pm 0.091) \times 10^{-2}$  & $0.950 \pm 0.052   $   & $0.0544 \pm 0.0031 $\\
$0.2035 \pm 0.0018 $               & $0.0224 \pm 0.0056  $ & $0.0172 \pm 0.0055 $\\
$0.41447 \pm 0.00034 $              & $0.0814 \pm 0.0018  $ & $0.0370 \pm 0.0011 $\\
\multicolumn{3}{|c|}{Ultraviolet}\\
$	8.337	\pm	0.042	  $  &  $  	0.042	\pm	0.011	  $  &  $  	0.75	\pm	0.15	  $  \\
$	9.47	\pm	0.31	  $  &  $  	0.028	\pm	0.049	  $  &  $  	1.1	\pm	1.1	  $  \\
$	10.37	\pm	0.23	  $  &  $  	0.044	\pm	0.083	  $  &  $  	1.3	\pm	1.4	  $  \\
$	11.65	\pm	0.29	  $  &  $  	0.06	\pm	0.12	  $  &  $  	1.7	\pm	1.7	  $  \\
$	13.00	\pm	0.38	  $  &  $  	0.08	\pm	0.20	  $  &  $  	1.8	\pm	2.2	  $  \\
$	14.27	\pm	0.34	  $  &  $  	0.11	\pm	0.26	  $  &  $  	2.2	\pm	2.4	  $  \\
$	15.94	\pm	0.48	  $  &  $  	0.09	\pm	0.21	  $  &  $  	2.8	\pm	3.0	  $  \\
$	18.01	\pm	0.45	  $  &  $  	0.091	\pm	0.077	  $  &  $  	3.41	\pm	0.87	  $  \\
$	21.031	\pm	0.092	  $  &  $  	0.101	\pm	0.022	  $  &  $  	6.71	\pm	0.72	  $  \\
$	29.50	\pm	0.70	  $  &  $  	0.150	\pm	0.057	  $  &  $  	20.7	\pm	4.1	  $  \\
$	50.2	\pm	4.7	  $  &  $  	0.076	\pm	0.035	  $  &  $  	50.5	\pm	6.2	  $  \\\hline
\end{tabular}
\caption{Fitting parameters for ice cold water ($T=  273.55\,\rm{K}$).}
\label{tbl:fit0}
\end{table}

\begin{figure}
    \centering
    \includegraphics[width=0.6\columnwidth]{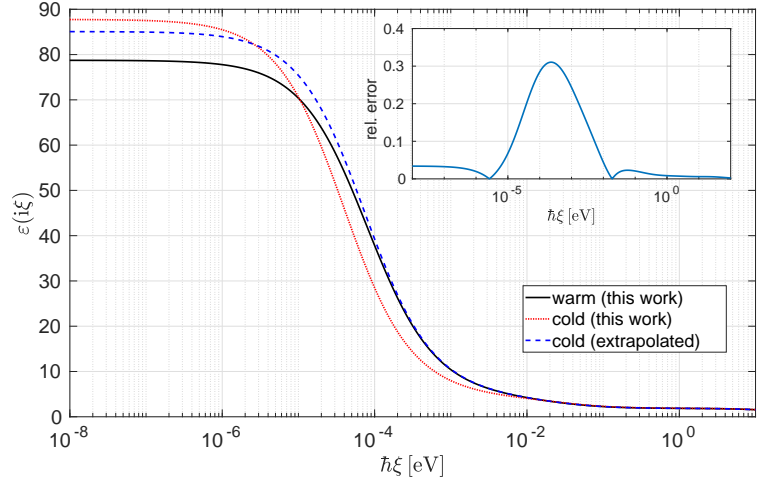}
    \caption{Dielectric functions of water at imaginary frequencies from this work for warm ($25^\circ$C) and ice-cold ($0.4^\circ$C) water, compared to the density-based extrapolation between temperatures proposed by Dagastine \textit{et al.}~\cite{Dagastine298Kwater}. The relative deviation is plotted as an inset.}
    \label{fig:water_warm_cold_extrapolate}
\end{figure}

Dagastine \textit{et al.}~\cite{Dagastine298Kwater} proposed that the temperature dependence of the dielectric function $\varepsilon(\mi\omega)$ at imaginary frequencies could be estimated via Clausius--Mossotti extrapolation using the temperature dependence of the density of water, with 
\begin{equation}
{\frac {\varepsilon(\mi \omega_n, T)-1} {\varepsilon(\mi \omega_n,T)+2}} \approx{\frac{\varrho(T)} {\varrho (T_{ref})}}{\frac {\varepsilon(\mi \omega_n,T_{ref})-1} {\varepsilon(\mi \omega_n,T_{ref})+2}}  \, ,
\end{equation}
We tested this extrapolation from warm water to ice-cold water, see Fig.~\ref{fig:water_warm_cold_extrapolate}, with \cite{CRC18} 
$\rho(25^\circ$C)= 0.9970470 g/cm$^{3}$ and $\rho(0.1^\circ$C)= 0.9998495 g/cm$^{3}$.  We find that the density-based extrapolation could be justified for low frequencies (radio-microwave $<10^{-5}$ eV) with relative error less than 5\%.  At microwave-IR frequencies ($10^{-5}$--$10^{-2}$ eV), however, the extrapolation performs poorly with relative error as high as 44\%, overestimating the true value. Superficially the extrapolation is again reasonable (error $<5$\%) in the near-IR , but only because the  difference between warm and ice-cold water is small at these higher frequencies. From these observations it would seem that density-based extrapolation is valid only in respect to the rotational polarisability of water, corresponding to the contribution of the Debye oscillators. Certainly, the Debye relaxation is determined by the number of hydrogen bonds broken during rotation~\cite{BUCHNER199957}, with Eyring-type temperature dependence $\tau(T) = (h/k_{\rm B} T) \mathrm{e}^{\Delta G/k_{\rm B}T}$ for some activation free energy $\Delta G(T)$.    But density-based extrapolation fails to capture the phonon contribution due to molecular vibrations in the IR spectrum.  Although we used a common real frequency UV data set for both temperatures when fitting,  our assumption that the UV spectrum is largely independent of temperature is corroborated by the small deviation between fitted and extrapolated ice-cold water at high frequencies.  

\section{Impact of the model on physical effects}

In order to illustrate the consequences of the new model for the dielectric function for water, we consider effects that include a wide spectral range. Dispersion forces provide an appropriate test case. We test  the interaction of air bubbles in water near specific materials chosen such that one water model predicts repulsion, while the other predicts attraction. Further, we consider the adsoprtion of  dissolved molecules at an water--air interface, and finally turn to the question of the freezing of water and premelting of ice. Consideration of these scenarios requires the transition to imaginary frequencies, $\varepsilon(\mi\xi)$, which is always a real quantity. Results    for the dielectric function of water are depicted in Fig.~\ref{fig:epsixi}.

\subsection{Air bubble near a dielectric interface}

As was already known to Lifshitz and co-workers~\cite{Dzya}, material combinations where an intervening medium (e.g. water) has a dielectric function lying  between those of two interacting surfaces leads to repulsive dispersion forces. Other material combinations where the dielectric function of the intervening medium is below or above both those of the interacting surfaces lead to attraction~\cite{PhysRevB.93.085434}. Based on this basic knowledge we propose a critical experimental test to check the validity of the different models for cold water dielectric functions. Due to the difference in magnitude for the dielectric functions  for imaginary frequencies as seen in Fig.~\ref{fig:epsixi} where the cold water model systematically has higher values than those reported by Elbaum and Schick~\cite{Elbaum273Kwater}, it should be possible to find a medium such that for a very large frequency range its dielectric function lies  between the two models for ice cold water. In that way the interaction between an air bubble in water near a surface would give opposite signs for the Hamaker--Lifshitz interaction if one water model is replaced with the other. Further tests could then be done to measure the Hamaker--Lifshitz interaction between a metallic surface (with optical density above the other materials), e.g. a gold nanotip in water near a certain material, where the opposite trends would be seen.

While the effect is so strongly dependent on the permittivity of water $\varepsilon_{\rm W}$, for the purpose of testing we model the behaviour of water using the  substrate with a simple permittivity $\varepsilon_{\rm X}$ modelled by one  single oscillator \cite{HOUGH19803}
\begin{equation}
    \varepsilon_{\rm X}(\mi\xi) = 1+ \frac{f}{1+(\xi/\omega_0)^2}.
    \label{dielmodelsolid}
\end{equation}
 The Hamaker constant changes sign depending on the water model,
\begin{equation}
    	H_{\rm AWX}=\frac{3k_{\rm B} T}{2}\sum_{n=0}^\infty {}'\frac{{\rm{Li}}_3[-r^{(\rm A)}(\mi\omega_n) r^{(\rm X)}(\mi\omega_n)]}{\varepsilon_{\rm W}(\mi\omega_n)}\,,
\label{eq:Hamaker}
\end{equation}
where the polylogarithmic function is given by
$\text{Li}_3(y)=4\int\limits_{0}^{\infty}\mathrm d x \, x^2\frac{y \mathrm e^{-2x}}{1-y \mathrm e^{-2x}}=\sum_{k=1}^\infty y^k/k^3$. The  reflection coefficients at the interfaces are $r^{(n)} =(\varepsilon_{n}-\varepsilon_{\rm W})/(\varepsilon_{n}+\varepsilon_{\rm W})$. The prime on the summation sign means that the $n=0$ mode is taken with half weight. 

\begin{figure}
    \centering
    \includegraphics[width=0.6\textwidth]{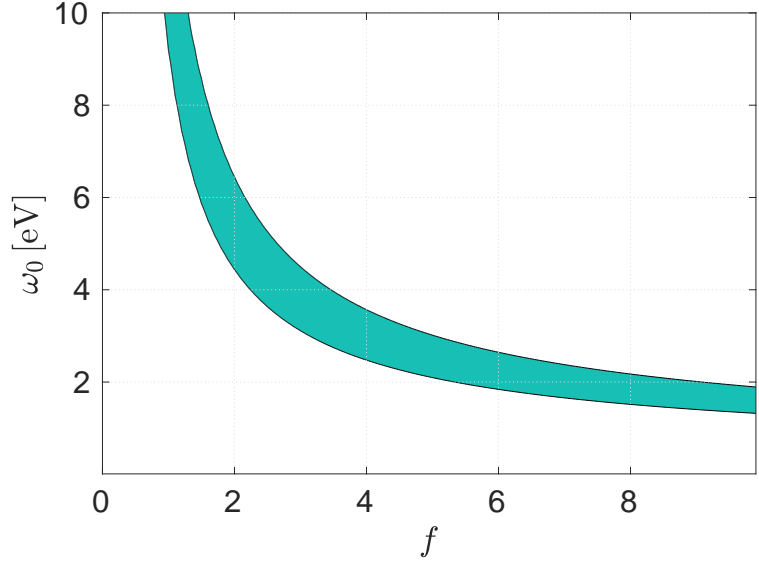}
    \caption{Contour plot marking the parameter space ($\omega_0$ and $f$) for a solid modelled by Eq.~(\ref{dielmodelsolid}) leading to the distinguishability between the presented model for ice-cold water at 273.55\,K and that from Elbaum and Schick~\cite{Elbaum273Kwater} for an air bubble in water near the surface. Here the coloured area marks the range where the parameters for the solid leads to different sign for the Hamaker interaction. The area above (below) leads to repulsion (attraction) for both water models.}
    \label{fig:my_label}
\end{figure}

As a guide for how to experimentally select a substrate with dielectric function between those of the two water models we start with a model substrate where the parameters $f$ and $\omega_0$ can be tuned such that there is a combination of parameters (the green coloured area in Fig.~\ref{fig:my_label}) such that   one water model gives repulsion and the other attraction. All substrates with parameters below the green coloured area give attraction for both water models (since the water models have a higher dielectric function than both substrate and bubble). Materials with parameters above the green coloured area lead to repulsion for bubbles in cold water near the substrate. These results are helpful when  identifying materials for experiments that can distinguish the more appropriate model for cold water.

\subsection{Casimir--Polder potential for particles dissolved in water}
Our second test deploys a non-retarded Casimir--Polder interaction acting between a molecule dissolved in water and a surface~\cite{Casimir48}. In general, this potential is described by an exchange of virtual photons~\cite{Buhmann12a}.
For a particle dissolved in a medium with permittivity $\varepsilon$ at finite temperature, the interaction energy at the interface of the medium with air (vacuum) in the non-retarded limit  takes the form~\cite{Buhmann12a,Buhmann12b,Friedrich}
\begin{equation}
U (z) \approx-{\frac {C_3} {z^3}}\,,\ C_3= \frac{k_{\rm B} T}{8\pi \varepsilon_0} \sum\limits_{n = 0}^{\infty}{'}  {\frac { \alpha^{\star}(\mi \omega_n)  }  {\varepsilon(\mi \omega_n) }} \left( \frac{1-\varepsilon(\mi \omega_n)}{1+\varepsilon(\mi \omega_n)} \right) \,,
\label{eq:CasimirPolder}
\end{equation}
where   $\omega_n=2 \pi k_{\rm B} T n/\hbar$ are the discrete Matsubara frequencies~\cite{Dzya}, $k_{\rm B}$ is the Boltzmann constant, and $\hbar$ is the reduced Planck constant. $\alpha^{\star}$ is the effective polarisability of the molecule in the medium. It can be directly seen that usually the resulting force is repulsive due to the negative Fresnel reflection coefficient describing the reflection from an optical dense to an optical thinner medium. With respect to the environmental medium surrounding the particles, we apply the real cavity model for local-field corrections~\cite{Johannes,Sambale07}, taking
\begin{equation}
\alpha^\star=\alpha^\star_C+\alpha\,
 \biggl(\frac{3\varepsilon}{2\varepsilon+1}\biggr)^2
 \frac{1}{1+2 \alpha^\star_C\alpha/(8\pi^2\varepsilon_0^2\varepsilon a_C^6)} \, ,
\label{eq:effectPol}
\end{equation}
with
\begin{equation}
\alpha^\star_C=4\pi\varepsilon_0\varepsilon a_C^3\,\frac{1-\varepsilon}
 {1+2\varepsilon}\, ,
\end{equation}
denoting the excess polarisability of the cavity (without the molecule).  $a_C$ is the radius of the cavity.

We apply the various water models to dissolved greenhouse gas molecules in water. The $C_3$ coefficients for the interaction with the air-water interface can be estimated by applying Eq.~(\ref{eq:CasimirPolder}) with Eq.~\ref{eq:effectPol}. 
The resulting $C_3$ coefficients are given  in Table~\ref{tbl:C3n_new}  with data for the polarisabilities and radii  taken from Ref.~\cite{Johannes}. These $C_3$-coefficients depend very sensitively on the water dielectric response model. This will influence predictions for how greenhouse gas molecules escape from melting ice~\cite{JohannesMelting2019}.

\begin{table}\centering
\begin{tabular}{|c|cccccc|cc|}\hline
 & \multicolumn{6}{c|}{$T=298.15 \, \rm {K}$} & \multicolumn{2}{c|}{$T=273.55 \, \rm {K}$}\\ 
Molecule & P\&W~\cite{PARSEGIAN1981285} & R\&L~\cite{ROTH1996637} & D.\textit{et al.}~\cite{Dagastine298Kwater} & W\&N~\cite{WANG201754} & W\&N corr. &\multicolumn{2}{c}{This work} & E\&S~\cite{Elbaum273Kwater}\\\hline
CH$_4$ & 52.7 & 51.1 & 106.3 & 423.2 & 34.4 & 15.0 & 94.8 & -45.6 \\ 
CO$_2$ & 172.5 & 170.2 & 235.0 & 685.1 & 144.4 & 115.2 & 224.6 & 31.6 \\ 
N$_2$O & 51.1 & 49.0 & 105.3 & 502.7 & 21.8 & 1.3 & 95.2 & -68.6 \\ 
O$_3$ & 100.3 & 98.2 & 154.3 & 564.7 & 71.0 & 48.1 & 145.1 & -25.0 \\ 
O$_2$ & 205.6 & 203.8 & 251.7 & 589.4 & 187.0 & 161.6 & 244.9 & 96.5 \\ 
N$_2$ & 173.9 & 172.1 & 221.6 & 558.3 & 155.0 & 130.6 & 213.9 & 66.1 \\ 
CO & 238.2 & 236.3 & 300.4 & 692.6 & 220.6 & 189.0 & 289.6 & 109.6 \\ 
NO$_2$ & 125.6 & 123.5 & 181.9 & 595.0 & 97.9 & 73.1 & 172.2 & -2.1 \\ 
H$_2$S & -102.2 & -103.6 & -35.8 & 273.1 & -119.9 & -134.0 & -53.7 & -194.2 \\
NO & 181.1 & 179.4 & 228.7 & 563.4 & 162.7 & 138.1 & 221.1 & 73.6 \\ \hline
\end{tabular}
\caption{$C_3$ dispersion coefficients [$\mu\mathrm{eV\, (nm)}^{3}$] for  dissolved particles in water interacting with the air--water interface.} \label{tbl:C3n_new}
\end{table}

 \subsection{Ice growth and premelting at the triple point of water}
In the absence of seeding particles, ice freezes from a water surface rather than from inside bulk water due to differences in density. This is the famous Archimedes principle at play via buoyancy force~\cite{Thiyam2018}. It is also well known that an ice surface both above and slightly below the freezing temperature of water has a thin film of melted water at its surface~\cite{Elbaum273Kwater}. Elbaum and Schick raised the important question whether  Lifshitz interactions could by themselves contribute sufficient energy for ice formation  at a water surface. They found, however, for a planar system with a thin ice film at the interface of water and vapour, that the short range interaction is attractive, suggesting a thin ice film would diminish and thus lead to  no surface freezing~\cite{Elbaum2}.  In other words, the Lifshitz energy was minimized when there was no ice formation on the water surface. In contrast, they found the remarkable result that ice tends to have a finite (few nanometres thick) premelted quasi-liquid water layer on its  surface at the triple point of water due to Lifshitz forces~\cite{Elbaum273Kwater}. In real systems surface charges, ions and impurities will increase the thickness of the premelted water film by orders of magnitude~\cite{Elbaum1993,Wilen,Wettlaufer,Thiyam2018}.

\begin{figure}
    \centering
    \includegraphics[width=0.6\textwidth]{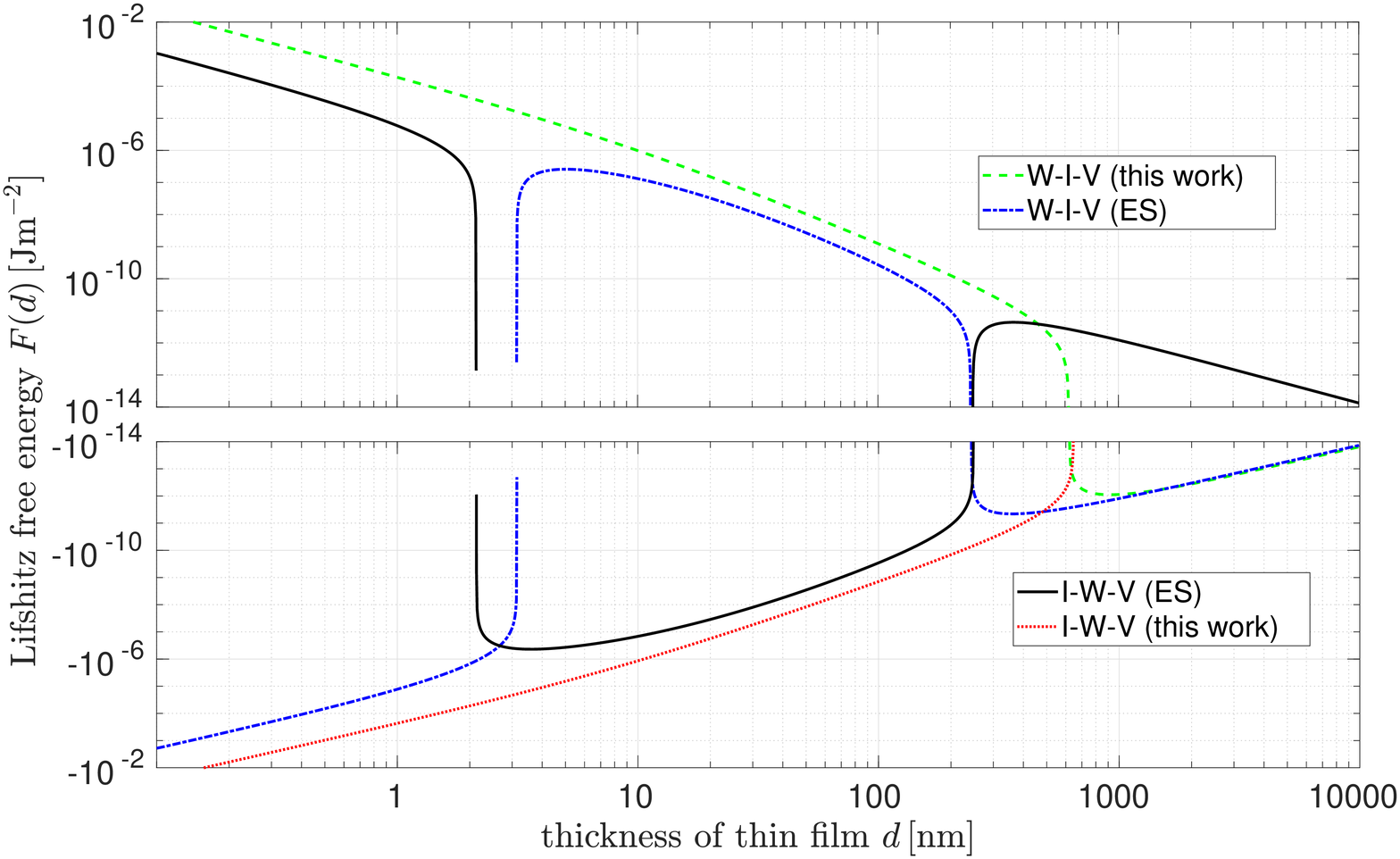}
    \caption{The interaction free energy for water--ice--vapour (W-I-V) and for ice--water--vapour (I-W-V) using Elbaum and Schick's model for dielectric function for ice. For water we use both Elbaum--Schick (ES) and the new model for water that we propose.}
    \label{fig:Icemeltingfreezing}
\end{figure}

In the context of the present work the question arises about  what will happen if our improved dielectric function for ice cold water is applied together with the available dielectric function for ice for systems at the triple point of water. In Fig.~\ref{fig:Icemeltingfreezing} we show the Lifshitz energies as a function of the intermediate layer thickness $d$ for ice--water--vapour and for water--ice--vapour using for ice the dielectric functions from Elbaum and Schick~\cite{Elbaum273Kwater} and for water Elbaum and Schick~\cite{Elbaum273Kwater} verses the new model for ice cold water proposed in this work. We reproduce the classic results from Elbaum and Schick that ice has a few nanometer thick premelted quasiliquid water layer on its interface between  water and vapour at the triple point for water~\cite{Elbaum273Kwater}. We also recover the result with no ice formation on a water--vapour interface~\cite{Elbaum2}. A very different story is found if we assume that the dielectric function for ice is correct and combine it with our dielectric function for cold water (one should note that our cold water corresponds to $T=273.55 \,\rm{K}$ rather than at the triple point). Here a thick, micron-sized, ice layer is predicted from Fig.~\ref{fig:Icemeltingfreezing} to grow on a water surface. What prevents the repulsion for water--ice--water system to lead to complete freezing is the attractive zero-frequency contribution in  the Lifshitz free energy, and the fact that other effects beyond dispersion forces are important.

As some  doubts can be raised about the water and ice models proposed by Elbaum and Schick~\cite{Elbaum273Kwater}, it is obvious  that further work is required. We stress that in order to have firm predictions for ice--water systems improved modelling of both water and ice based on accurate optical data from both a large frequency range and for many different temperatures and pressures are essential.

One may  be concerned by the fact that use of  the new water model suggests no ice premelting. This is troublesome since this is well established to occur. However, as was demonstrated by Wettlaufer, impurities and salt effects can never be ignored. They  are independent of the water and ice models for dielectric functions, and are clearly important effects behind ice premelting~\cite{Elbaum1993,Wilen,Wettlaufer}. Here, one  should also recall the recent predictions from Thiyam \textit{et al.}~\cite{Thiyam2018} that ionic forces, due to self-consistently set-up pH-dependent charges at the ice--water interface in a salt concentration,  will often be dominant and, typically, be repulsive. 

Lastly, an unavoidable  question is how much the water dielectric function will change very close to the phase transition from liquid to solid state, leading to effects for the dielectric function for quasiliquid water at the triple point not accounted for by considering cold liquid water only.

\section{Conclusion}

A thorough study of available optical data for room temperature water, and ice cold water, has been carried out. This enabled us to develop an improved model for the real and imaginary parts of the dielectric functions for water and, as an important side product, the  dielectric functions for imaginary frequencies. The model has been tested against available models in the literature. Sensitive effects where the accuracy of water dielectric models is important involve gas molecules dissolved in water near surfaces. Surprising  effects are also observed when an improved model for ice cold water near the triple point is included in the Lifshitz theory for intermolecular forces suggesting ice formation  at water--air interfaces. 
One main point is that since the predicted results are so sensitive, realistic predictions for ice formation and melting can only be arrived at once water and ice dielectric functions  have both been very accurately modelled. The ultimate aim with our work is to inspire more experiments to be performed, thus providing a larger set of input data for the modelling  of liquid and solid water for a wide range of frequencies, as well as for different temperatures and pressures.

\section*{Acknowledgements}
We especially thank Richard Buchner and Hisashi Hayashi for fruitful discussions and sharing data (Ref.~\cite{BUCHNER199957} and Ref.~\cite{Hayashi15} respectively). We acknowledge support from the Research Council of Norway (Project  250346). We gratefully acknowledge support by the German Research Council (grant BU1803/3-1, SYB and JF).

\bibliographystyle{unsrt}  
\bibliography{sample}  






\end{document}